**Understanding Magnetic Phase Coexistence in Ru$_2$Mn$_{1-x}$Fe$_x$Sn Heusler Alloys:**
**A Neutron Scattering, Thermodynamic, and Phenomenological Analysis**

Eric McCalla[1,2], Emily E. Levin[3], Jason E. Douglas[3,4] , John G. Barker[4], Matthias Frontzek[5],

Wei Tian[5], Rafael M. Fernandes[6], Ram Seshadri[3] and Chris Leighton[1*]

[1]*Department of Chemical Engineering and Materials Science, University of Minnesota,*

*Minneapolis, Minnesota 55455, USA*

[2]*Department of Chemistry, McGill University,*

*Montréal, Québec, Canada, H3A 0B8*

[3]*Materials Department and Materials Research Laboratory,*

*University of California Santa Barbara, California 93106, USA*

[4]*National Institute of Standards and Technology,*

*100 Bureau Drive, Gaithersburg, Maryland 20899 USA*

[5]*Neutron Scattering Division, Oak Ridge National Laboratory,*

*Oak Ridge, Tennessee 37830, USA*

[6]*School of Physics and Astronomy, University of Minnesota,*

*Minneapolis, Minnesota 55455, USA*

*Corresponding author: [leighton@umn.edu](mailto:leighton@umn.edu)

**ABSTRACT:** The random substitutional solid solution between the antiferromagnetic (AFM) full-Heusler alloy Ru$_2$MnSn and the ferromagnetic (FM) full-Heusler alloy Ru$_2$FeSn provides a rare opportunity to study FM-AFM phase competition in a near-lattice-matched, cubic system, with full solubility. At intermediate $x$ in Ru$_2$Mn$_{1-x}$Fe$_x$Sn this system displays suppressed magnetic ordering temperatures, spatially coexisting FM and AFM order, and strong coercivity enhancement, despite rigorous chemical homogeneity. Here, we construct the most detailed temperature- and $x$-dependent understanding of the magnetic phase competition and coexistence in this system to date, combining wide-temperature-range neutron diffraction and small-angle neutron scattering with magnetometry and specific heat measurements on thoroughly characterized polycrystals. A complete magnetic phase diagram is generated, showing FM-AFM coexistence between $x \approx 0.30$ and $x \approx 0.70$. Important new insight is gained from the extracted length scales for magnetic phase coexistence (25-100 nm), the relative magnetic volume fractions and ordering temperatures, in addition to remarkable $x$-dependent trends in magnetic and electronic contributions to specific heat. An unusual feature in the magnetic phase diagram (an intermediate FM phase) is also shown to arise from an extrinsic effect related to a minor Ru-rich secondary phase. The established magnetic phase diagram is then discussed with the aid of phenomenological modeling, clarifying the nature of the mesoscale phase coexistence with respect to the understanding of disordered Heisenberg models.

PHYSH:      Research areas:        Magnetism; magnetic interactions, magnetic order

            Physical systems:      Magnetic systems; alloys; Heusler alloys

            Techniques:            Neutron scattering, magnetization measurements, specific

                                   heat measurements

## I. INTRODUCTION

The chemically-ordered intermetallics known as Heusler alloys have grown to encompass a large family, with broad potential applications [1,2]. These alloys crystallize in cubic full-Heusler ($X_2YZ$) and half-Heusler ($XYZ$) variants, incorporating a variety of X and Y transition metals (*e.g.*, Mn, Fe, Co, Ru) and Z main group elements (*e.g.*, Al, Si, Ga, Ge, In, Sn, Sb) [1,2]. Magnetic examples from this alloy class provide perfect illustrations of their diverse functionalities, which include: ferromagnetism with nonmagnetic X, Y, Z; high Curie temperature ($T_C$) and saturation magnetization ($M_S$); half-metallic or highly spin-polarized character; spin-gapless semiconducting behavior; magnetocaloric response; and exciting topological characteristics [1-4]. Particularly extensively studied in this context are NiMnSb and the $Co_2MnZ$ and $Ni_2MnZ$ families, although interest has been widespread [1-4]. A pervasive strategy in the investigation of such materials is to study quaternary solid solution or deliberately off-stoichiometric versions, such as $Co_2FeGe_{1-x}Ga_x$, $Ni_2Mn_{1+x}Sn_{1-x}$, *etc.*, which enable composition-based tuning of lattice parameter, spin-polarized electronic structure, topological band structure, and so on [1-4].

In fundamental magnetism, a particularly attractive prospect with quaternary solid solution and/or off-stoichiometric Heusler alloys is the identification of model systems to study the long-standing problem of ferromagnetic-antiferromagnetic (FM-AFM) phase competition. Prominent examples include heavily studied $Ni_2Mn_{1+x}Sn_{1-x}$, $Ni_2Mn_{1+x}In_{1-x}$, and $Ni_{2-x}Co_xMn_{1+y}Sn_{1-y}$, where AFM interactions (*e.g.*, due to Mn-Mn interactions generated by $Mn_{Sn}$ substitution) are controllably introduced into a FM Heusler matrix (*e.g.*, $Ni_2MnSn$) [5-17]. The impacts of the resulting FM-AFM phase competition are fascinating, encompassing phase separation into short-range FM clusters in AFM or non-magnetically-ordered matrices, resulting superparamagnetism in bulk solids, and exchange bias and coercivity enhancement due to naturally-formed FM/AFM interfaces [5-17]. Such effects also interplay with martensitic phase transformations, generating magnetic-field-induced transformations, magnetic shape memory effects, magnetocaloric and barocaloric phenomena, *etc.* [1,2,5-17].

Quaternary solid solution Heuslers, such as $Cu_{1-x}Ni_xMnSb$ [18] and $Ru_2Mn_{1-x}Fe_xSn$ [19-21], have also attracted attention as more idealized systems for the study of FM-AFM phase competition. In $Ru_2Mn_{1-x}Fe_xSn$, this is because $Ru_2FeSn$ is FM with $T_C \approx 500$ K, while $Ru_2MnSn$ is AFM with Néel temperature $T_N \approx 300$ K [19-21]. Compositional tuning on a single site then enables the study of FM-AFM phase competition with complete solubility in $Ru_2Mn_{1-x}Fe_xSn$, while maintaining cubic structure, with only a ~0.02 Å variation in lattice parameter [19-21]. Such complete solubility between FM and AFM end-points in a cubic system with modest structural variations is rare. In recent work, polycrystalline $Ru_2Mn_{1-x}Fe_xSn$ was synthesized, quenched to avoid chemical phase separation, and proven chemically homogeneous over broad length scales [19]. This homogeneity was established *via* powder X-ray diffraction (PXRD), neutron powder diffraction (NPD), pair distribution function (PDF) analysis of neutron scattering data, and scanning electron microscopy (SEM) and transmission electron microscopy (TEM) with energy dispersive X-ray spectroscopy (EDS) [19]. Temperature ($T$)-dependent magnetometry then revealed suppression of $T_N$ as $x$ increases from zero (*i.e.*, as Fe is alloyed into $Ru_2MnSn$), and rapid suppression of $T_C$ as $x$ decreases from unity (*i.e.*, as Mn is alloyed into $Ru_2FeSn$) [19]. Critically, low $T$ NPD at $x = 0.50$ then revealed both FM *and* AFM reflections, with no canted or ferrimagnetic state, suggesting *spatial coexistence* of FM and AFM phases, despite the chemical homogeneity [19]. In the FM-AFM coexistence region, a striking low $T$ coercivity enhancement was discovered, *i.e.*, magnetic hardening, ascribed to interfacial FM/AFM interactions.

This work on $Ru_2Mn_{1-x}Fe_xSn$ quickly stimulated theoretical studies. In 2017, density functional theory (DFT) computations shed significant light by establishing the origin of FM in $Ru_2FeSn$ and AFM in $Ru_2MnSn$ [20]. The moment in such systems is strongly confined to Fe and Mn, with the FM or AFM arising from a subtle balance between Sn-mediated AFM superexchange and the itinerant electron FM RKKY (Ruderman-Kittel-Kasuya-Yoshida) interaction [20]. These calculations also highlighted a tendency to chemical phase separation [20], no doubt frustrated by quenching, as in other Heusler-based systems [22-24]. A tendency to form (111)-oriented stripes and short-range-ordered clusters of Fe- and Mn-rich FM and AFM phases was uncovered [20]. In 2019, Decolvenaere *et al*. built on this to advance a mixed-basis chemical and magnetic cluster expansion method, to which Monte Carlo simulations were applied, both to equilibrated and quenched structures [21]. This enabled semi-quantitative reproduction of the available magnetic phase diagram, accurately describing $M_S(x)$ [21]. The Monte Carlo simulations also provided snapshots of the separation into spatially coexisting short-range-ordered FM and AFM regions, providing much insight [21].

While the above represents substantial progress with $Ru_2Mn_{1-x}Fe_xSn$, challenges and questions remain. It would be desirable, for example, to extend the limited neutron scattering data (which are at low $T$ only, at $x = 0.00$, 0.50, and 1.00 [19]) to a complete study *vs*. $x$ and $T$, to construct a full magnetic phase diagram. The critical issue of the *length scales* over which the FM-AFM phase coexistence occurs is also poorly understood, due to limited application of experimental probes with appropriate spatial resolution. With respect to experimental probes in general, magnetometry, some NPD, Mössbauer spectroscopy, local structure methods, and basic structural and chemical characterization have been applied [19], but thermodynamic and transport studies are absent. This is despite the utility of the latter for probing magnetically inhomogeneous systems.

In light of the above, we present here a detailed wide-$T$-range neutron scattering study of competing FM and AFM order in $Ru_2Mn_{1-x}Fe_xSn$, spanning $x = 0.00$, 0.25, 0.40, 0.50, 0.60, and 1.00, using both NPD and small-angle neutron scattering (SANS). This is combined with magnetometry, and, importantly, specific heat measurements, along with analysis of $T$-dependent lattice parameters, to provide the most comprehensive understanding to date of the $T$- and $x$-dependent FM-AFM phase competition and coexistence in $Ru_2Mn_{1-x}Fe_xSn$. Nano- to meso-scale FM-AFM phase coexistence is thus pinned down to $0.30 < x < 0.70$, where we extract detailed information on $T$-dependent AFM and FM order parameters, $T_N$ and $T_C$, AFM and FM volume fractions, and phase coexistence length scales. Specific heat, in addition to providing the $x$-dependent Debye temperature, reveals signatures of both the FM phase and AFM-FM phase coexistence, as well as an enhancement of the Sommerfeld coefficient around $x = 0.50$. An unusual feature in the phase diagram (an intermediate FM phase) is also elucidated as arising from an extrinsic effect, related to a Ru-rich secondary phase. Finally, guided by phenomenological modeling, the deduced experimental magnetic phase diagram is appropriately placed in the context of the theoretical understanding of Heisenberg models for disordered magnets.

## II. EXPERIMENTAL DETAILS

Polycrystalline $Ru_2Mn_{1-x}Fe_xSn$ samples were prepared *via* solid-state synthesis from Mn, Fe, Ru and Sn, then quenched in ice water, as described earlier [19]. The exact samples used by Douglas *et al.* [19] were employed in our magnetometry, SANS, and specific heat measurements. These samples were previously characterized by PXRD (with Rietveld refinement), SEM, TEM, SEM- and TEM-based scanning EDS, low $T$ NPD (at $x = 0.00$, 0.50, and 1.00 only), and neutron PDF analysis [19]. As discussed in Section I, chemical phase separation was ruled out at all probed length scales [19]. For the more extensive NPD in the current paper, new, higher mass (~3-5 g) samples were prepared *via* similar methods, resulting in similar PXRD, room temperature lattice parameters, *etc.* A previously identified Ru-rich hexagonally-close-packed (HCP) minor secondary phase (5% mole fraction) [19] was also detected in these NPD samples.

Magnetometry was done in a Quantum Design PPMS vibrating sample magnetometer (VSM) with a high temperature oven, from 5-700 K, in applied magnetic field ($H$) to 50 kOe. Heat capacity measurements were also performed in a PPMS (1.8 to 380 K, zero field), using relaxation calorimetry. 2% temperature pulses were used, 3 measurements were averaged at each $T$, the thermal coupling factor never fell below ~95%, and the ratio of sample to addenda heat capacity was maintained above 1.5 [25]. NPD was done at Oak Ridge National Laboratory on the HB-1A (FIE-TAX) and WAND instruments of the High Flux Isotope Reactor (HFIR). On HB-1A, a fixed incident energy of 14.6 meV was used, employing a double pyrolytic graphite (PG) monochromator system. Two highly-oriented PG filters were placed after each monochromator to reduce higher-order contamination of the incident beam. Collimator settings of open-40'-sample-40'-80' were used. On WAND (HB-2C) a focusing 113-Ge monochromator with an incident wavelength of 1.486 Å was used, in tandem with a 1D position sensitive detector; the 113-Ge reflection has no wavelength contamination. The resulting resolution $dQ/Q$ (where $Q$ is the scattering wavevector) is approximately 0.02 in the used $Q$ range. On both instruments, a 15-700 K temperature interval was explored, using high temperature closed-cycle refrigerators. Samples were mounted in sealed V cans in helium exchange gas. SANS measurements were done on the NG7 and NGB 30 m beamlines at the NIST Center for Neutron Research, using sample-detector distances of 3 and 12 m to span a $Q$ range of 0.005-0.16 Å$^{-1}$. Data were taken from 5 to 650 K, in a high temperature closed-cycle refrigerator.

## III. RESULTS AND ANALYSIS

In the interests of clarity, in Section III.A we first present the deduced experimental magnetic phase diagram of $Ru_2Mn_{1-x}Fe_xSn$, augmenting our findings with previously reported data. Section III.B then discusses the $T$-dependent NPD measurements used to track the order parameters and ordering temperatures upon which the phase diagram is based, along with additional information on length scales associated with the magnetic ordering. Complementary magnetometry measurements are then presented in Section III.C. SANS data providing further insight into FM order and associated length scales are provided in Section III.D, followed by $x$- and $T$-dependent specific heat measurements in Section III.E. The latter are then connected to $T$-dependent lattice parameter anomalies in Section III.F.

## III.A MAGNETIC PHASE DIAGRAM

Fig. 1(a) depicts the $Ru_2Mn_{1-x}Fe_xSn$ magnetic phase diagram deduced in this work. The $T_N$ and $T_C$ values shown derive from the NPD measurements presented in Section III.B, corroborated by magnetometry and SANS in Sections III.C and III.D. As discussed in the Introduction, AFM order with $T_N \approx 300$ K occurs at $x = 0.00$, $i.e.$, in $Ru_2MnSn$. As $x$ increases, $T_N$ gradually decreases to ~250 K at $x \approx 0.30$, above which a drastic change occurs. Specifically, FM order is also now detected, coexisting with AFM. At $x = 0.40$, 0.50, and 0.60, for example, FM ordering emerges first on cooling (at a $T_C$ that increases rapidly with $x$), followed by AFM ordering at a lower $T_N$ that decreases rapidly with $x$, reaching ~120 K at $x = 0.60$; $T_C$ and $T_N$ thus appear uncoupled. Increasing $x$ beyond ~0.70 then results in all signatures of AFM order being lost in neutron and magnetometry measurements, $i.e.$, phase-pure FM. (Note that while compositions between 0.60 and 1.00 are not studied here, prior work suggests phase-pure FM at $x = 0.75$ [19], meaning that the region of FM-AFM phase coexistence ends between 0.60 and 0.75; we thus take $x \approx 0.7$ as an estimate). This phase-pure FM behavior persists to $x = 1.00$ (Ru$_2$FeSn), at which point $T_C$ reaches

500 K. Based on Fig. 1(a), the low $T$ (ground state) magnetic phase behavior with increasing $x$ can thus be characterized as phase-pure AFM to $x \approx 0.30$, and phase-pure FM beyond $x \approx 0.70$,

bracketing a substantial range ($0.30 < x < 0.70$) over which FM and AFM order coexist. Crucially

(see Section III.B), NPD data at $0.30 < x < 0.70$ indicate: (*i*) no effect of the onset of AFM order

(at $T_N$) on the development of FM order, and (*ii*) no evidence for a new magnetically-ordered state such as a canted AFM or a ferrimagnet. *Spatially coexisting* FM and AFM order is instead implicated. The only other phase in Fig. 1(a) [aside from paramagnetic (PM) at high $T$] is the one labeled FM*, which will be discussed in Sections III.D and III.F, where it is concluded that this is related to the minor Ru-rich secondary phase. As returned to below (Section IV), broadly similar phase diagrams have been reported in other quaternary Heuslers, particularly Cu$_{1-x}$Ni$_x$MnSb [18].

Reinforcing the phase diagram in Fig. 1(a), shown in Fig. 1(b) are the low $T$ (15 K) relative NPD

scattering intensities (normalized to their maximum values) from the AFM (red) and FM (blue)

phases. Consistent with the end of the phase-pure AFM region in Fig. 1(a), the relative AFM

intensity, a proxy for the AFM phase fraction, is seen to drop from 1.0 between $x = 0.25$ and 0.40,

remaining finite to $x \approx 0.70$, *i.e.*, throughout the AFM-FM coexistence region. Correspondingly,

the relative FM intensity is zero to between 0.25 and 0.40, above which it grows at the expense of

the AFM phase, saturating at 1.0 (phase-pure FM) at around 0.70. AFM and FM order parameters

thus coexist between approximately $x = 0.30$ and 0.70. (Again, while compositions between 0.60

and 1.00 are not studied here, prior work suggests phase-pure FM at $x = 0.75$ [19], meaning that the region of FM-AFM phase coexistence ends between 0.60 and 0.75; we thus take $x \approx 0.7$ as an estimate). Also shown for completeness in Fig. 1(b) are the 4 K coercivity values from Ref. 19, showing the magnetic hardening due to interfacial AFM/FM coupling. Interestingly, this peaks at $x = 0.40$, not at the composition where FM and AFM volume fractions cross, but at lower $x$, where small volume fractions of FM phase are embedded in an AFM matrix.

## III.B NEUTRON DIFFRACTION

Fig. 2(a) first shows an example NPD pattern ($x = 0.50$) in the low $Q$ region, where magnetic information can be obtained. Consistent with prior work at this composition [19], on cooling to low $T$ both FM *and* AFM orderings are evident, through the growth of the FM 1 1 1 reflection, and the emergence of the AFM 1/2 -3/2 1/2 reflection (among others). This AFM order corresponds to AFM coupling between (111) planes of parallel spins [19], as in other Full-Heuslers such as $Ru_2MnSb$ and $Ru_2MnGe$ [26]. Most importantly, and again consistent with prior work [19], possibilities such as canted AFM and ferrimagnetism were found inconsistent with these NPD data. Specifically, all low $T$ NPD patterns in the $0.30 < x < 0.70$ regime could not be fit with a single magnetic wavevector, implying spatial coexistence of FM and AFM order, as opposed to a new magnetic phase.

This conclusion is further supported by detailed $x$- and $T$-dependent measurements of the FM and AFM order, using the intense 1 1 1 and 1/2 -3/2 1/2 NPD reflections; these are the measurements used to determine the phase behavior in Fig. 1. Fig. 2(b) first shows the $T$ dependence of the relative FM intensity (normalized to the low $T$, $x = 1.00$ value), which reveals clear Curie points followed by order-parameter-like growth for $x = 1.00, 0.60, 0.50$, and $0.40$; the low FM intensity in the $x = 0.40$ case is magnified in the inset. The low $T$ saturation value of the relative FM intensity is seen to gradually drop with decreasing $x$ (as in Fig. 1(b)), mirroring the decrease in $T_C$ (as in Fig. 1(a)). At $x = 0.25$ and below, no FM order was detected by NPD (as in Fig. 1(a)), consistent with phase-pure AFM. Fig. 2(c) then shows equivalent data for the relative AFM intensity (normalized to the low $T$, $x = 0.00$ value), which reveals clear Néel points followed by order-parameter-like growth for $x = 0.00$, 0.25, 0.40, 0.50, and 0.60. Accompanying the gradual decrease in $T_N$ (as in Fig. 1(a)), the low $T$ saturation value of the relative AFM intensity drops as $x$ is increased above 0.25 (as in Fig. 1(b)), becoming undetectable above $x = 0.60$, consistent with phase-pure FM. As shown in Figs. 1(a,b), these low $T$ NPD data thus support phase-pure AFM to $x \approx 0.30$, phase-pure FM beyond $x \approx 0.70$, and FM-AFM coexistence between ~0.30 and ~0.70. Critically, and consistent with the conclusion of spatially coexisting FM and AFM as opposed to a canted AFM or ferrimagnet, $T_C$ and $T_N$ appear essentially uncoupled. The onset of AFM order at a $T_N$ lower than $T_C$ (see Fig. 1(a), for example) has no apparent impact on the growth of the FM order parameter. The solid lines in Figs. 2(b,c) are in fact fits to squared mean-field order parameters (see caption for details), confirming mean-field-like behavior even in the FM-AFM coexistence regime. At $x = 0.50$, for example, FM order sets in at $T_C \approx 280$ K (Fig. 2(b)), the AFM ordering at $T_N \approx 180$ K (Fig. 2(c)) having no impact on the FM order parameter (Fig. 2(b)).

The important issue of the length scales associated with FM and AFM order is addressed in Figs. 2(d,e). Plotted here are the $T$ dependences of the Scherrer lengths ($\Lambda$) extracted from the 1 1 1 FM (Fig. 2(d)) and 1/2 -3/2 1/2 AFM (Fig. 2(e)) reflections, *i.e.*, the lengths calculated by applying the Scherrer equation to the peak full-widths at half-maximum. These values were corrected for the instrumental broadenings, which were determined from reference data on Si powder, HB-1A (FIE-TAX) having slightly lower broadening than WAND. Starting with the AFM order probed in Fig. 2(e), at $x = 0.00$ and 0.25 the low $T$ value of $\Lambda$ saturates at about 65 nm, which is practically indistinguishable from full long-range order on the instruments used (as approximately denoted by the horizontal dashed line). As illustrated by the $x = 0$ data in Fig. 2(d), where no FM is present, this is also the length scale extracted from the nuclear peak widths of these samples. We thus conclude that the AFM order at $x = 0.00$ and 0.25 is long-range within instrumental resolution.

The $x = 0.40$, 0.50, and 0.60 behavior in Fig. 2(e), however, is different. In this phase coexistence regime the low $T$ length scales saturate at 25-50 nm, consistent with shorter-range AFM. In all cases, the extracted AFM length scales decrease as $T \rightarrow T_N^-$, as expected. Moving to FM order (Fig. 2(d)), the fact that FM NPD peaks coincide with nuclear peaks, rather than emerging from the background as for AFM peaks, poses additional challenges. Nevertheless, the low $T$ values of $\Lambda$ are again minimum around $x = 0.50$ and 0.60, in the phase coexistence regime, pointing to shorter-range FM order, on 35-50 nm scales. We thus conclude that the coexisting low $T$ FM and AFM orders at $0.30 < x < 0.70$ occur with correlation lengths of a few 10s of nm, *i.e.*, at the nano-to-meso-scale. For the FM compositions, more details on these length scales are provided by SANS (Section III.D).

## III.C MAGNETOMETRY

$T$-dependent magnetometry measurements providing complementary insight to NPD are provided in Fig. 3, at representative $x = 1.00$ (a), 0.60 (b), 0.50 (c), and 0.40 (d). The temperature dependence of the magnetization ($M$) is shown at $H = 50$, 200, 500, 5000, and 50000 Oe, spanning a wider $H$ range than prior work. At $x = 1.00$, the behavior is unremarkable, the sharp feature at the FM $T_C$ of ~530 K simply broadening as $H$ is increased. (Note that minor variations in ordering temperatures occur in comparison to larger mass samples used for NPD, ascribed to small differences in preparation and composition). As $x$ is decreased to 0.60, 0.50, and 0.40, however (Figs. 3(b-d), note the different $T$ scale to Fig. 3(a)), FM-AFM coexistence kicks in, both $T_C$ *and* $T_N$ becoming apparent in $M(T)$. At the highest $H$, $M(T)$ in Figs. 3(b-d) is essentially featureless below $T_C$, consistent with the mean-field FM order parameter growth in Fig. 2(a). As $H$ is decreased, however, $T_N$ becomes progressively apparent *via* features in $M(T)$, and conspicuous bifurcation of field-cooled (solid lines) and zero-field-cooled curves (dashed lines). As expected, the high $H$, low $T$ saturation magnetization also drops with decreasing $x$, tracking the relative FM NPD intensity in Figs. 2(a) and 1(b). Magnetometry is thus in good agreement with Figs. 1 and 2, particularly the phase-pure FM behavior at $x > 0.70$, and the FM-AFM coexistence at lower $x$.

### III.D SMALL-ANGLE NEUTRON SCATTERING (SANS)

SANS is a powerful probe of FM ordering and inhomogeneity (particularly at nano- and meso-scales) [27] and was thus applied here at $x = 1.00$, 0.60, 0.50, and 0.40, $i.e.$, where FM was detected by NPD and magnetometry. (As a low scattering wavevector ($Q$) technique, SANS is specifically sensitive to FM ($i.e.$, $Q = 0$) fluctuations and order [27]). Shown first in Fig. 4 are representative SANS cross-section ($d\Sigma/d\Omega$) $vs.$ $Q$ plots for $x = 1.00$ (top panels), 0.60 (middle panels), and 0.40 (bottom panels), at high (right panels), intermediate (middle panels), and low $T$ (left panels). At $x = 1.00$, the high $T$ ($i.e.$, 600 K) $d\Sigma/d\Omega(Q)$ in Fig. 4(c) is composed of two contributions: a low $Q$ contribution with linear behavior on this $\log_{10}$-$\log_{10}$ plot ($i.e.$, a power-law), and a high $Q$ contribution with a slower roll-off to the highest $Q$. As illustrated by the dashed lines, these two contributions are well described by Porod and Lorentzian terms, $i.e.$, $d\Sigma/d\Omega = (d\Sigma/d\Omega)_P/Q^n$ and $d\Sigma/d\Omega = (d\Sigma/d\Omega)_L/(Q^2 + 1/\xi^2)$, respectively, where $(d\Sigma/d\Omega)_P$ and $(d\Sigma/d\Omega)_L$ parameterize the strength of the Porod and Lorentzian scattering, $n$ is the Porod exponent, and $\xi$ is the Ornstein-Zernike magnetic correlation length [27]. Our data were best fit with $n = 4.15$ (blue dashed lines in Fig. 4), close to the classic $n = 4$ exponent for Porod scattering from three-dimensional objects of size $d$ with smooth surfaces, in the $Q >> 2\pi/d$ limit [27]. As is typical, we ascribe this to scattering from microstructural features such as grains and grain boundaries above $T_C$, and from FM domains and domain walls below $T_C$ [27]. In these unpolarized measurements, the low $Q$ Porod intensity thus saturates at a $T$-independent level at $T > T_C$, but grows below $T_C$. This can be seen by comparing Figs. 4(a-c), where $d\Sigma/d\Omega$ at the lowest $Q$ grows significantly on cooling, as shown more clearly below. The Lorentzian contribution, on the other hand (red dashed line), captures the short-range FM spin fluctuations that grow as $T \rightarrow T_C^+$ at a second-order paramagnetic-to-FM phase transition [27]. This contribution is thus strong in Fig. 4(c), at $T = 600$ K (= $1.13T_C$), but diminishes rapidly on cooling (*e.g.*, Fig. 4(a)).

Considering Figs. 4(a-c) together, at $x = 1.00$ the observed behavior is thus fairly typical: the Porod contribution grows on cooling below $T_C$ due to long-range FM, while the Lorentzian scattering is strong around and above $T_C$ but diminishes on cooling. As exemplified by Fig. 4(b), however, an additional contribution emerges at intermediate $T$ (480 K in this case). This consists of a small but distinct hump at intermediate $Q$ of ~0.015 Å$^{-1}$, which can be captured (green dashed line) by a Gaussian peak, *i.e.*, $d\Sigma/d\Omega = (d\Sigma/d\Omega)_G \exp[-(Q-Q_G)^2/(2\Delta_G^2)]$, where $(d\Sigma/d\Omega)_G$ is the peak intensity, $Q_G$ is the peak position, and $\Delta_G$ is the peak width. This is a somewhat atypical feature, not of obvious origin, although similar behavior has been found in off-stoichiometric magnetic Heusler alloys, due to nanoscopic FM clusters in paramagnetic, AFM, or even FM matrices [13,17,27]. The origin in the current case will be clarified below, in Sections III.E and F. Most important for now, with these three contributions (low $Q$ Porod, intermediate $Q$ Gaussian peak, and high $Q$ Lorentzian), all data can be fit to

$$\frac{d\Sigma}{d\Omega}(Q,T) = \frac{\left(\frac{d\Sigma}{d\Omega}\right)_P(T)}{Q^n} + \frac{\left(\frac{d\Sigma}{d\Omega}\right)_L(T)}{Q^2 + \left(\frac{1}{\xi}\right)^2} + \left(\frac{d\Sigma}{d\Omega}\right)_G(T) exp\left(\frac{-[Q-Q_G(T)]^2}{2\Delta_G(T)^2}\right) \qquad (1),$$

resulting in the solid lines through the data in Fig. 4, and the Porod, Gaussian, and Lorentzian contributions shown in dashed blue, green, and red. At $x = 0.60$ (where $T_C \approx 300$ K) the Lorentzian scattering is again strong near $T_C$ (Fig. 4(f)), then diminishes on cooling as the Porod scattering from long-range FM grows, with a small Gaussian hump again showing up at intermediate $T$ (Fig. 4(e)). The same qualitative trends are then repeated for $x = 0.40$ (Figs. 4(g-i)).

The overall $T$ dependence for $x = 1.00, 0.60, 0.50$, and $0.40$ is shown more clearly in Fig. 5, simply by plotting the low and high $Q$ scattering cross-sections *vs*. $T$. Scattering wavevectors of 0.006 and 0.114 Å$^{-1}$ were chosen, *i.e.*, at the lower and upper ends of the probed range, where Porod and Lorentzian contributions dominate, respectively. As shown in Figs. 5(a,b,) the minimum $d\Sigma/d\Omega$ at any $T$ was subtracted here, employing the standard approach to isolate $T$-dependent magnetic scattering in unpolarized SANS. As in Figs. 2(a) and 3(a), the behavior for $x = 1.00$ is that of an archetypal long-range-ordered FM, the low $Q$ magnetic Porod scattering (Fig. 5(a)) turning on at $T_C$ then growing monotonically [27]. Correspondingly, in Fig. 5(b) the high $Q$ (Lorentzian-dominated) magnetic scattering grows as $T \rightarrow T_C^+$ then falls quickly below $T_C$, vanishing at low $T$; this is a classic "critical scattering" peak [27]. Decreasing $x$ into the FM-AFM coexistence regime at $x = 0.60$ and 0.50 then leads to suppressed $T_C$, as expected, with lower magnetic Porod intensity (Fig. 5(a)), as well as critical scattering peaks (Fig. 5(b)). As deduced from NPD, clear FM order thus persists in this composition regime. Regarding the length scales of this FM order, the Porod scattering clearly includes a significant magnetic contribution at $x = 0.60$ and 0.50 (Fig. 5(a)), which extends to the minimum $Q$ in Figs. 4(d,g) of 0.005 Å$^{-1}$. This corresponds to scattering from FM domains of size above $2\pi/0.005$ Å$^{-1} \approx 100$ nm, reasonably consistent with the 35-55 nm from NPD (Fig. 2(c)).

Finally, at $x = 0.40$, the behavior in Fig. 5 changes. At this composition, at which FM order was weak in NPD, no order-parameter-like growth is seen in Fig. 5(a) and no critical scattering peak is seen in Fig. 5(b). The primary feature is instead weak monotonic growth of the high $Q$ magnetic scattering intensity on cooling (Fig. 5(b)), indicating short-range FM spin fluctuations but no long-range FM order. This sample is thus very close to the onset of phase-pure AFM and the end of the FM-AFM coexistence regime, the only semblance of FM being short-ranged. The weak long-range

FM order detected at $x = 0.40$ by NPD (Fig. 2(a), inset), and the only short-range FM detected by SANS (Fig. 5(b)) are likely reconcilable *via* the minor sample-to-sample compositional and magnetic property variations already noted (*e.g.*, with respect to the small variations in $T_C$ between NPD and SANS samples with $x = 1.0$, $0.60$, and $0.50$).

More quantitative SANS analysis is provided in Fig. 6, which shows the $T$ dependence of the parameters extracted from fits of equation 1 to $d\Sigma/d\Omega(Q)$ at all $T$ and $x$. Note here that while the number of fitting parameters in (1) is significant, in the low $Q$ region the Porod term is entirely dominant and in the high $Q$ region the Lorentzian term is entirely dominant, leading to high confidence in the extracted parameters. Shown first in Fig. 6(a) is $(d\Sigma/d\Omega)_P$, which, consistent with Fig. 5(a), shows monotonic increases on cooling below $T_C$ for $x = 1.00$, $0.60$, and $0.50$. The solid lines are in fact fits to squared mean-field order parameters, confirming quantitative agreement with NPD (Fig. 2(a)). No such behavior occurs for $x = 0.40$, however, consistent with Figs. 5(a,b). Correspondingly, $(d\Sigma/d\Omega)_L$ (Fig. 6(b)) is substantial above $T_C$ for $x = 1.00$, $0.60$, and $0.50$ (due to short-range FM spin fluctuations), falling rapidly below $T_C$. At $x = 0.40$, consistent with Fig. 5(b), no critical scattering occurs, $(d\Sigma/d\Omega)_L$ instead growing weakly on cooling to the lowest $T$, indicating short-range FM spin fluctuations only. As expected, the extracted $\xi(T)$ for $x = 1.00$, $0.60$, and $0.50$ (Fig. 6(c)) then exhibits power-law divergence as $T \rightarrow T_C^+$, *i.e.*, $\xi = \xi_o / (T/T_C - 1)^\nu$, where $\xi_o$ is a constant and $\nu$ is a critical exponent [27,28]. The solid line fits in Fig. 6(c) yield $\nu = 0.63$, $0.78$, and $0.73$ for $x = 0.5$, $0.6$ and $1.0$, respectively. These are close to the three-dimensional Heisenberg and Ising FM exponents, confirming typical behavior [27,28]; we make no attempt at deeper quantitative analysis, as the required detailed, small-$T$-spacing data were not taken. Due to the very low scattering intensity, no such $\xi$ values could be extracted at low $T$ for $x = 0.40$.

Figs. 6(d,e) then plot the parameters related to the intermediate $Q$ Gaussian scattering. Given the modest intensity of this hump (Figs. 4(b,e,h)) one challenge here is that it is difficult, likely futile, to separate $Q_G$, the peak position, from $\Delta_G$, the width. We instead fixed $Q_G = 0.013$ Å$^{-1}$ based on preliminary fits, and left $(d\Sigma/d\Omega)_G$ and $\Delta_G$ as the only parameters. The behavior of $(d\Sigma/d\Omega)_G(T)$ in Fig. 6(d) (which is robust with respect to different fitting approaches) is remarkable, revealing that this anomalous intermediate $Q$ scattering occurs only over a finite $T$ window for $x = 1.00$, 0.60, and 0.50. Specifically, at $x = 1.00$, $(d\Sigma/d\Omega)_G$ turns on at $T_C$, grows rapidly, but then diminishes below ~500 K, becoming undetectable at ~340 K (as in Fig. 4(a)). Similar behavior occurs for $x = 0.60$ and 0.50, but shifted to lower $T$, with lower intensity. In the phase diagram in Fig. 1(a) we thus define a new region, FM$^*$, between $T_C$ and the temperature $T^*$ where $(d\Sigma/d\Omega)_G$ vanishes. Fig. 6(e) plots $\Delta_G(T)$, the width of the intermediate $Q$ Gaussian hump, which, as already noted (see Figs. 4(b,e,h)) is challenging to separate from $Q_G$, the peak position. The extracted $\Delta_G$ values appear to have systematic $T$ dependence (decreasing on cooling for each $x$), falling in the approximate range $0.002 – 0.035$ Å$^{-1}$. The length scales extracted from the positions and widths of the anomalous intermediate $Q$ peak are thus in the 10s to 100s of nm range. The physical meaning of these is returned to below, when the FM$^*$ region is clarified in Sections III. E and F.

### III.E SPECIFIC HEAT

Specific heat ($C_P$) provides a direct, powerful probe of magnetic ordering and has been extensively utilized to probe magnetic phase coexistence and phase separation, and was thus also applied here. Fig. 7(a) shows the standard analysis of low $T$ (<10 K in this case) $C_P(T)$ in metals, plotting $C_P/T$ vs. $T^2$ to probe for behavior of the form $C_P(T) = \gamma T + \beta T^3$. The first term here describes electronic excitations, where $\gamma$, the Sommerfeld coefficient, is given by $\gamma = \pi^2 k_B^2 D(E_F)/3$, with $k_B$

Boltzmann's constant and $D(E_F)$ the density of states at the Fermi level [29]. The second term describes lattice excitations, which, in the low $T$ limit of the Debye model, yield $\beta = 234Nk_B/\Theta_D^3$, where $N$ is the number of atoms per mole and $\Theta_D$ is the Debye temperature [29]. As can be seen from Fig. 7(a), while approximately linear behavior with a positive intercept is found in Ru$_2$Mn$_{1-}$$_x$Fe$_x$Sn at most $x$, deviations do arise. The most obvious occurs for $x = 0$ (Ru$_2$MnSn), which appears to show divergence as $T \rightarrow 0$ on this $C_P/T$ $vs.$ $T^2$ plot. Such behavior is not uncommon in low $T$ specific heat, often indicating Schottky anomalies linked to energy level spacings that arise from myriad factors, including paramagnetism, crystal field splittings, nuclear hyperfine contributions in magnetically-ordered systems, $etc.$ [30-37]. While this obscures further low $T$ analysis of $C_P(T)$, this occurs at only one composition ($x = 0.00$), and would require further work at lower $T$ to elucidate. We thus leave this as a topic for future work. At higher $x$, particularly 0.40 and above, while the intercept in Fig. 7(a) indicates finite $\gamma$, as expected in these metals, small deviations from linear remain. Close inspection ($e.g.$, at $x = 0.40$, $0.50$, $0.60$) reveals that these are $downward$ deviations from linear on cooling, indicating $C_P(T)$ contributions weaker than $T^3$. Given that spin-waves in phase-pure long-range-ordered FMs often give a $C_P$ contribution $\propto T^{3/2}$ [$e.g.$, 32-34], and that $T^2$ specific heat has been observed in various magnetically-phase-separated systems [32-36], the data were fit to

$$C_P(T) = \gamma T + \beta T^3 + B_1 T^2 + B_2 T^{3/2} \qquad (2),$$

where $B_1$ and $B_2$ are pre-factors of the $T^2$ and $T^{3/2}$ terms. The result is excellent fits at all $x$ other than 0.00, yielding the $\Theta_D$, $\gamma$, $B_1$, and $B_2$ shown $vs.$ $x$ in Figs. 7(b-d).

The $\Theta_D(x)$ data in Fig. 7(b) are fairly unremarkable. (Note here that the $x = 0.00$ value shown was determined from higher $T$ than in Fig. 7(a)). Only a modest variation in $\Theta_D$ occurs across the entire



phase diagram (~40-50 K), the increase upon substitution of Fe for Mn being unsurprising, as are the overall $\Theta_D$ values, which can be compared to 310, 330, 460, 505, and 550 K in $Ni_2Mn_{1.4}Sn_{0.6}$ [38], $Cu_2MnAl$ [39], $Ru_2VGa$ [40], $Cu_2MnSn$ [39], and $Ru_2VAl$ [40] full-Heuslers, respectively. The behavior of $\gamma$, however, is more interesting. Specifically, $\gamma$ exhibits a broad maximum around $x = 0.5$, rising from ~9 mJ $mol^{-1}$ $K^{-2}$ for $x = 1.00$ (fairly typical for a metallic full-Heusler alloy), to ~17 mJ $mol^{-1}$ $K^{-2}$ around $x = 0.50$ (quite large for a metallic full Heusler). Applying $\gamma = \pi^2 k_B^2 D(E_F)/3$ yields $D(E_F)$ from ~3.5 up to ~7.0 states/eV/formula unit. Interestingly, the DFT calculations of Decolvenaere et al. [20] also suggest non-monotonic $D(E_F)$ vs. $x$ in $Ru_2Mn_{1-x}Fe_xSn$, rising from 3.07 states/eV/formula unit at $x = 0.00$, to 3.59 states/eV/formula unit at $x = 0.33$, then falling to 1.86 states/eV/formula unit at $x = 1.00$. This arises from complex $x$-dependent changes in the spin-resolved $D(E_F)$, reflecting the delicate magnetic phase competition. Nevertheless, the overall non-monotonic trend in Fig. 7(c) is at least qualitatively consistent with first-principles results. More quantitatively, the enhancement of the measured $\gamma$ over DFT is close to 2, independent of $x$. This is well within the realm of typical mass enhancement factors due to electron-phonon interactions, electronic correlations, etc. We note that magnetic behavior such as spin-glass freezing is known to produce additional $T$-linear contributions to $C_P(T)$ [41], but this is unlikely here given the FM-AFM coexistence as opposed to glassy magnetism. As discussed below, the magnetism in $Ru_2Mn_{1-x}Fe_xSn$ manifests in other terms in equation 2.

Moving to these other terms, we first see in Fig. 7(d) that non-zero $B_2$ was only required to fit the $C_P(T)$ data for $x = 1.00$, i.e., in the phase-pure long-range FM state, where $B_1$ is negligible. Regarding statistical significance, a free fit with $C_P(T) = \gamma T + \beta T^3 + cT^n$ (with $c$ and $n$ constants) significantly improved $\chi^2$, resulting in $n = 1.48$, i.e., very close to 3/2. Therefore, while deviations



from linearity for $x = 1.00$ in Fig. 7(a) are small, the additional $T$ dependence is conclusively of $B_2T^{3/2}$ form, as expected for a long-range-ordered FM [*e.g.*, 32-34]. The $B_2$ of 0.4 mJ mol$^{-1}$ K$^{-2.5}$, while not out of bounds in comparison to other FMs, is small, so it is unsurprising that this $C_P$ contribution is not detected at lower $x$, particularly given the maxima in $\gamma(x)$ and $B_1(x)$. Effects of long-range FM order on $C_P(T)$ are also known to fall off quickly as phase-pure FM is disrupted [36]. Moving to $B_1(x)$, the striking feature is the prominent peak at intermediate $x$, the composition range over which $B_1 \neq 0$ corresponding exactly to the FM-AFM coexistence regime in Fig. 1; FM-AFM phase coexistence is thus clearly manifested in $C_P(T)$. This $T^2$ contribution to $C_P$ in magnetically-phase-separated systems has a substantial history, since Woodfield *et al.* suggested that coexisting FM-like and AFM-like excitations in an A-type AFM could generate $T^2$ [32]. This $T^2$ contribution was subsequently found in various magnetically-phase-separated systems, including manganites and cobaltites, suggesting that it could be a general signature of nanoscopic FM regions embedded in non-FM matrices [33-36]. The detection of this $C_P$ contribution here, in a metal alloy, is thus of high interest, particularly given the peak in $B_1$ at $x = 0.5$, *i.e.*, the exact composition at which the FM and AFM volume fractions phases cross (Fig. 1(b)). In terms of statistical significance, consistent with the visible downward curvature on cooling for $x = 0.40$, 0.50, and 0.60 in Fig. 7(a), reasonable fits could not be obtained without finite $B_1$ in equation 2. Free fits to $C_P(T) = \gamma T + \beta T^3 + cT^n$ also yield $n = 1.8$ to 2.0 in all cases, leading to high confidence in the conclusion of $T^2$ specific heat related to FM-AFM coexistence.

Further insight is provided by the higher $T$ behavior of $C_P(T)$. The $x = 0.5$ case is shown as illustrative in Fig. 8(a), revealing qualitatively typical form, approaching $3R$ at high $T$. Notably, no lambda anomaly is seen in Fig. 8(a) (where $T_N \approx 180$ K), as is also the case at 0.40 and 0.60. Although at first surprising, it should be noted, as alluded to above, that lambda anomalies at



second-order magnetic phase transitions have been reported to fall off quickly as phase-pure long-range FM or AFM is disrupted by doping into a regime of magnetic phase separation [36]. Outside of the FM-AFM coexistence regime, where $T_C$ and $T_N$ should be visible in $C_P(T)$, they fall outside our measurement range (*e.g.*, $T_C = 500$ K at $x = 1.00$), or in a range where the vacuum grease used to affix samples produces spurious features (*e.g.*, $T_N = 300$ K at $x = 0.00$). Nevertheless, one aspect of the phase behavior in Fig. 1 *is* detected in $C_P(T)$, namely $T^*$. As illustrated in Figs. 8(b-d), evidence exists (certainly for $x = 1.00$, but also 0.60 and 0.50) for $C_P(T)$ anomalies very close to the $T^*(x)$ from SANS. For $x = 1.00$ a clear step in $C_P(T)$ occurs at $T^*$ (Fig. 8(d)), becoming weaker at lower $x$, although a more subtle slope change persists (Figs. 8(b,c)). The $T^*$ values deduced from Figs. 8(b-d) are the open red points in Fig. 1(a), agreeing closely with $T^*$ from SANS (open green points, from Fig. 6(d)). The anomalous intermediate $T$ range FM$^*$ region in Fig. 1 is thus detected not only by SANS, but also $C_P(T)$. Inspired by Figs. 8(b-d), the NPD data of Section III.B were further analyzed to extract $T$-dependent lattice parameters in the FM phase, searching for an anomaly at $T^*$.

### III.F TEMPERATURE-DEPENDENT STRUCTURAL MEASUREMENTS

Shown in Fig. 9(a) are cubic lattice parameter (*a*) *vs.* $T$ data for $x = 0.40$, 0.50, 0.60, and 1.00, obtained from Pawley fits of NPD data. As expected, the room temperature $a$ decreases slightly with $x$, and thermal expansion occurs. The latter was analyzed by fitting $a(T)$ in Fig. 9(a) using the Grüneisen approximation for anharmonic phonon potentials combined with Einstein lattice dynamics, *i.e.*, $a(T) = a_0 \left\{ 1 + \frac{\alpha_0 T_E}{2} \left[ \coth\left( \frac{T_E}{2T} \right) - 1 \right] \right\}$, where $a_o$ is the $T = 0$ lattice parameter, $T_E$ is the Einstein temperature, and $\alpha_o$ is the linear thermal expansion coefficient at $T >> T_E$ [42]. The resulting fits (Fig. 9(a)) are generally good, particularly for $x = 0.40$, 0.50, and 0.60, revealing no detectable anomalies at $T_C$, $T_N$, or $T^*$. This is shown more quantitatively for $x = 0.50$ in the inset,



where the deviation between data and fit ($\Delta a$) is plotted *vs. T*, confirming no systematic deviations. At $x = 1.00$, however, deviations are visible in Fig. 9(a), the inset showing negative deviations (*i.e.*, experimental $a$ values below the fit) in broad ranges between ~100 and 350 K, and ~350 and 500 K. A fairly well-defined maximum in $\Delta a$ thus occurs at ~350 K, close to $T^*$ from SANS and specific heat (see Figs. 1(a), 6(d), and 8(d)). It is plausible that weaker anomalies at $T^*$ for $x = 0.40$, 0.50, and 0.60 are not detected simply due to the significant curvature in $a(T)$ at their respective $T^*$ (~110-150 K); the $x = 1.00$ composition is instead in the linear thermal expansion regime at its $T^* \approx 350$ K.

Further insight was obtained by analyzing the *T*-dependent lattice parameters of the previously mentioned Ru-rich minority phase. In prior work this phase was hypothesized, based primarily on Rietveld refinement of NPD data, to be a Ru-rich HCP Ru-Mn-Fe solid solution, likely forming coherently within the cubic Heusler matrix, and thus being strained [19]. Fig. 9(b) shows the *c*-axis lattice parameter of this HCP secondary phase ($c_{hcp}$) for $x = 1.00$, which, in addition to thermal expansion, reveals a clear anomaly at ~350 K, *i.e.*, $T^*$ at this $x$ (see Figs. 1(a), 6(d), 8(d), and 9(a) inset). The subtle feature at $T^*$ in $a(T)$ in the Heusler primary phase is thus conspicuous in the $c_{hcp}(T)$ of the HCP Ru-rich secondary phase, supporting the conjecture of structural coherence between the phases, and substantial strain. Considering this in light of Figs. 6(d), 8(b-d), and 9(a,b) clarifies the origin of the FM* phase and associated $T^*$ in Fig. 1(a). Specifically, the intermediate $Q$ hump in the SANS data in Fig. 4 indicates some form of inhomogeneity in the FM-ordered regions in $Ru_2Mn_{1-x}Fe_xSn$, which Fig. 6(d) establishes to occur only at $T^* < T < T_C$, on length scales (Fig. 6(e)) of 10s - 100s of nm. Figs. 8(b-d) and 9(a,b) further establish a subtle structural transition in the $Ru_2Mn_{1-x}Fe_xSn$ primary phase, strongly strain-coupled to the Ru-rich HCP secondary phase. Based on Fig. 6(d), this transition apparently leads to negligible magnetization contrast with the



FM Heusler primary phase at $T < T^*$, whereas this contrast is clearly visible at $T^* < T < T_C$. While the precise origin of this behavior is difficult to pin down, one possibility is FM ordering of the secondary Ru-Mn-Fe phase. At high $T$ (*i.e.*, $T^* < T < T_C$) the magnetic inhomogeneity detected by SANS in the FM phase of the $Ru_2Mn_{1-x}Fe_xSn$ would thus arise due to PM secondary phase regions dispersed in the FM Heusler matrix, the onset of FM in the secondary phase regions (at $T < T^*$) then decreasing the magnetization contrast and suppressing the intermediate $Q$ hump. Accompanying strain-coupled magnetostructural anomalies could then produce the behaviors in Figs. 9(b), 9(a, inset), and 8(b-d). Most importantly, regardless of the exact origin, the data of Fig. 9 establish $T^*$ as linked to the Ru-rich secondary phase, meaning that it is *not* fundamental to the phase behavior of $Ru_2Mn_{1-x}Fe_xSn$. We thus leave the FM* phase on Fig. 1(a) for completeness, but emphasize that is very likely extrinsic.

## IV. THEORETICAL DISCUSSION

With the aim to shed further light on the deduced experimental phase diagram (Fig. 1(a)), and better place it in the context of current understanding of magnetic phase competition in disordered magnets, we consider a general phenomenological model for competing FM and AFM states. Denoting the FM vector order parameter $\boldsymbol{M}_F$ and the AFM equivalent $\boldsymbol{M}_A$, the Landau free-energy expansion is (see, *e.g.*, Refs. [43, 44]):

$$F = \left(\frac{a_F}{2} M_F^2 + \frac{u_F}{4} M_F^4\right) + \left(\frac{a_A}{2} M_A^2 + \frac{u_A}{4} M_A^4\right) + \left[\frac{\gamma}{2} M_F^2 M_A^2 + \frac{w}{2} (\boldsymbol{M}_F \cdot \boldsymbol{M}_A)^2\right] \qquad (3),$$

where, $a_F \propto T - T_C$, $a_A \propto T - T_N$, and the other coefficients are quartic Landau parameters. We also assume $u_A, u_F > 0$ to ensure the pure FM and AFM transitions are of second-order nature. The transition temperatures $T_C$ and $T_N$ are then functions of a tuning parameter – in our case, the concentration $x$ of Fe – and cross at a multi-critical point. Depending on the relationship between the quartic Landau coefficients [43,44], the system may then display either a new thermodynamic mixed phase, where both order parameters coexist microscopically at all atomic sites, or a phase where FM and AFM orders coexist on a mesoscopic scale *without* forming a new phase, *i.e.*, in distinct regions of the sample. In the former case, the multi-critical point is tetracritical; inside the magnetically-ordered phase, as one moves along the phase diagram, there is a second-order transition from the AFM phase to the mixed phase, and then another second-order transition from the mixed phase to the FM phase. In the latter case, on the other hand, the multi-critical point is bicritical and there is a first-order transition between the AFM and FM phases, with mesoscopic AFM-FM coexistence as a consequence.

The mean-field "phase diagram" describing this behavior is shown in Fig. 10. Applying the standard criterion for tetracritical *vs.* bicritical behavior [45], the system is in a mixed phase when $-\sqrt{u_F u_A} < \gamma + \min(0, w) < \sqrt{u_F u_A}$, and displays mesoscopic coexistence otherwise. The nature of the mixed phase also depends on the sign of $w$: when $w > 0$, the AFM and FM order parameters coexist on the atomic scale and are mutually perpendicular, giving rise to a canted spin structure. When $w < 0$, however, the order parameters are parallel, resulting in a ferrimagnetic spin structure. Both structures are depicted schematically in Fig. 10. The experimental results reported here strongly support the scenario of mesoscopic AFM-FM phase coexistence, placing the Landau parameters in the regime corresponding to the yellow shaded areas in Fig. 10. First, as shown in Fig. 1(b), the AFM and FM volume fractions seem to simply grow at the expense of one another *vs. x*, suggesting little, if any, spatial overlap between phases. Second, Figs. 2(a-b) show that the squared FM order parameter $M_F^2$ is insensitive to the onset of AFM order at a lower $T$. This can be compared to known cases of competing orders coexisting on the atomic scale, such as AFM and superconductivity in iron-based materials, where the onset of the latter strongly suppresses the former [46]. Third, there is no signature of non-collinear or non-uniform (from site to site) magnetization in the various experimental probes employed here, ruling out a canted or ferrimagnetic spin structure.

Of course, such phenomenological analysis does not explain *why* the Landau parameters are such that the multi-critical point is bicritical in our case. For that, microscopic models are needed to determine those parameters. For $Ru_2Mn_{1-x}Fe_xSn$, the concentration $x$, which tunes the system from AFM to FM, has two effects: tuning the delicate balance between AFM and FM interactions and adding disorder ($Fe_{Mn}$ substitutions). The latter is commonly captured in nearest-neighbor Heisenberg models by introducing either site or bond disorder. For bond disorder, one often assumes a distribution of FM and AFM bonds across the system [44]. The expectation from analyses of the closely related Sherrington-Kirkpatrick model is that a spin-glass phase will then form between the FM and AFM phases (for a review, see [47]). No signatures of spin-glass effects appear in our experimental data, however. One can then also consider site-disorder. While there are different ways of implementing this, let us focus on the case where each site randomly has either a FM or AFM atomic species [43]. As discussed in Ref. [48], where Monte Carlo simulations of this model were performed, if there is no coupling between the two types of sites, and if the lattice is bipartite, the system displays two independent percolation-driven AFM and FM transitions, resulting in a so-called decoupled tetracritical point. The tetracritical point remains stable upon inclusion of coupling between the two types of sites, even in the frustrated regime [48]. While these results apply directly to simple and body-centered cubic lattices (SC and BCC, respectively), the situation for the non-bipartite face-centered cubic (FCC) lattice, most relevant here, is less clear. The frustration intrinsic to the FCC lattice does not prevent the onset of AFM order, which is expected to be collinear in the clean limit [49]. However, the Monte Carlo simulations of Ref. [48] did not probe this more involved type of frustrated AFM order. While it is possible that the frustration introduced by the FCC lattice results in an intermediate spin-glass phase, it is also conceivable that, similar to the cases of the BCC and SC lattices, and to the general results of Ref. [43], the multi-critical point remains tetracritical.

These effective Heisenberg models thus do not seem to entirely capture the phase diagram of $Ru_2Mn_{1-x}Fe_xSn$. In Ref. [21], however, a cluster-expansion Hamiltonian approach was recently developed to model this specific quaternary Heusler system, drawing on results from first-principles calculations. The key insight with this approach is to include not only the effects of magnetic disorder, but also chemical disorder. In the quenched case, a phase diagram remarkably similar to Fig. 1(a) was obtained, suggestive of a bicritical point. AFM-FM phase coexistence at very short length scales was in fact observed in the Monte Carlo simulations, with a wide $x$ range over which AFM (FM) clusters form in an FM (AFM) matrix, potentially explaining the exchange hardening in Fig. 1(b). Based on this, it is therefore highly likely that chemical disorder plays a key role in shaping the phase diagram of $Ru_2Mn_{1-x}Fe_xSn$. The observed AFM-FM phase coexistence can thus be at least qualitatively reconciled with established models, inclusion of quenched chemical disorder leading to more quantitative agreement.

## V. SUMMARY

In short, we have presented an NPD, SANS, magnetometry, specific heat, and structural-based analysis of the magnetic phase behavior of the $Ru_2Mn_{1-x}Fe_xSn$ system, culminating in a detailed magnetic phase diagram. Aside from an anomalous FM* phase associated with a subtle extrinsic effect related to a minority phase, the phase diagram essentially reveals phase-pure AFM to $x \approx 0.30$, phase-pure FM above $x \approx 0.70$, and a substantial regime of FM-AFM nano- to meso-scale coexistence between, despite chemical homogeneity. In this coexistence regime, the development of FM order below $T_C$ is essentially unperturbed by the onset of AFM order at a lower $T_N$, resulting, at low $T$, in intertwined FM and AFM order on 10s of nm length scales. $T_C$ is apparent in magnetometry, NPD and SANS, $T_N$ is apparent in magnetometry and NPD, the relevant length scales can be determined from NPD and SANS, and the low $T$ FM-AFM coexistence is found to be reflected in a $T^2$ contribution to specific heat. Comparison with theoretical models suggests that chemical disorder is likely key to stabilizing a bicritical point in the phase diagram (and thus mesoscopic FM-AFM phase coexistence), as opposed to the tetracritical point generally expected from Heisenberg models with AFM and FM site-disorder, or to the spin-glass phase typical of the Heisenberg model with AFM and FM bond-disorder. In totality, these findings thus substantially improve the understanding of the magnetic phase coexistence in this model system, thereby improving the overall understanding of the enduring topic of magnetic phase competition in compositionally-tuned systems.

**ACKNOWLEDGMENTS:** This work was supported primarily by the US Department of Energy through the University of Minnesota Center for Quantum Materials under DE-SC-0016371. EM acknowledges financial support from the Natural Sciences and Engineering Research Council of Canada. The work at UC Santa Barbara was supported by the National Science Foundation (NSF) Materials Research Science and Engineering Center (MRSEC) under DMR-1720256 (IRG-1). Part of this research used resources at the High Flux Isotope Reactor, a DOE Office of Science User Facility operated by the Oak Ridge National Laboratory. We thank Elizabeth Decolvenaere and Dominic Ryan for useful conversations.

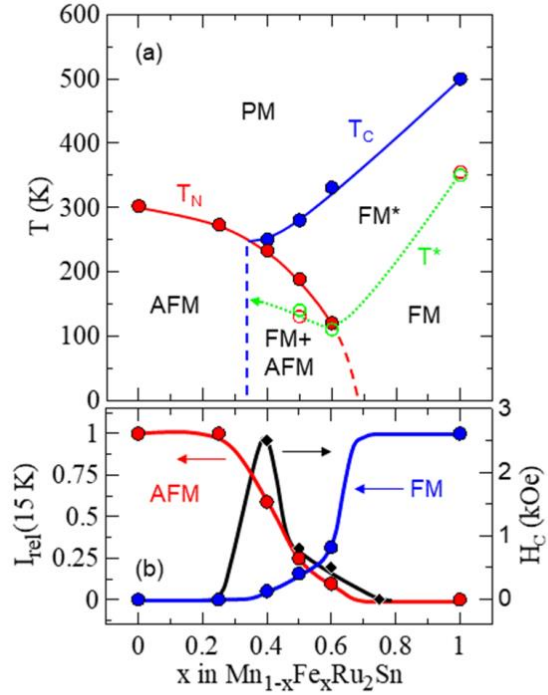

**Figure 1:** (a) $Ru_2Mn_{1-x}Fe_xSn$ magnetic phase diagram deduced from this work. The red, blue, and green points (data) and lines (guides to the eye) are the Néel temperature ($T_N$), Curie temperature ($T_C$), and $T^*$ (defined in the text). $T_N$ and $T_C$ were determined from neutron diffraction, and $T^*$ from SANS (open green points) and heat capacity (open red points). The labeled phases are PM (paramagnet), AFM (antiferromagnet), FM (ferromagnet) and FM$^*$ (as discussed in the text). The blue and red dashed lines illustrate the region over which FM-AFM phase coexistence is deduced. (b) Normalized neutron diffraction intensities ($I_{rel}$) of the FM 1 1 1 (blue) and AFM 3/2 1/2 1/2 (red) reflections at 15 K. Solid lines are guides to the eye. As noted in the text, while compositions between 0.60 and 1.00 were not studied here, prior work [19] suggests phase-pure FM at $x = 0.75$, meaning that FM-AFM phase coexistence ends between 0.60 and 0.75; we thus depict the red (blue) line in (b) to reach zero (unity) at $x \approx 0.7$ (where the FM-AFM coexistence region ends in (a)). The black points and line (corresponding to the right axis) are the 4 K coercivity ($H_c$) from Ref. 19.

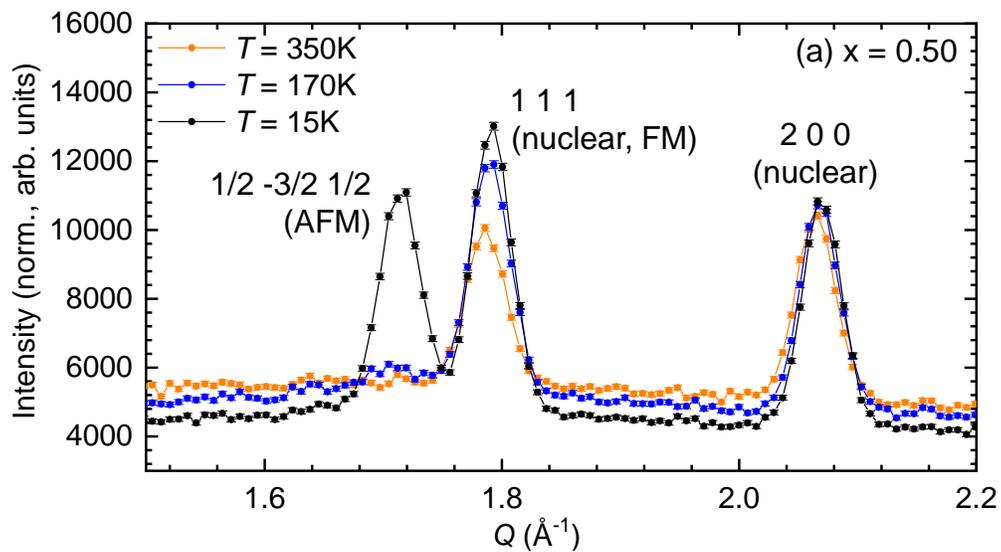

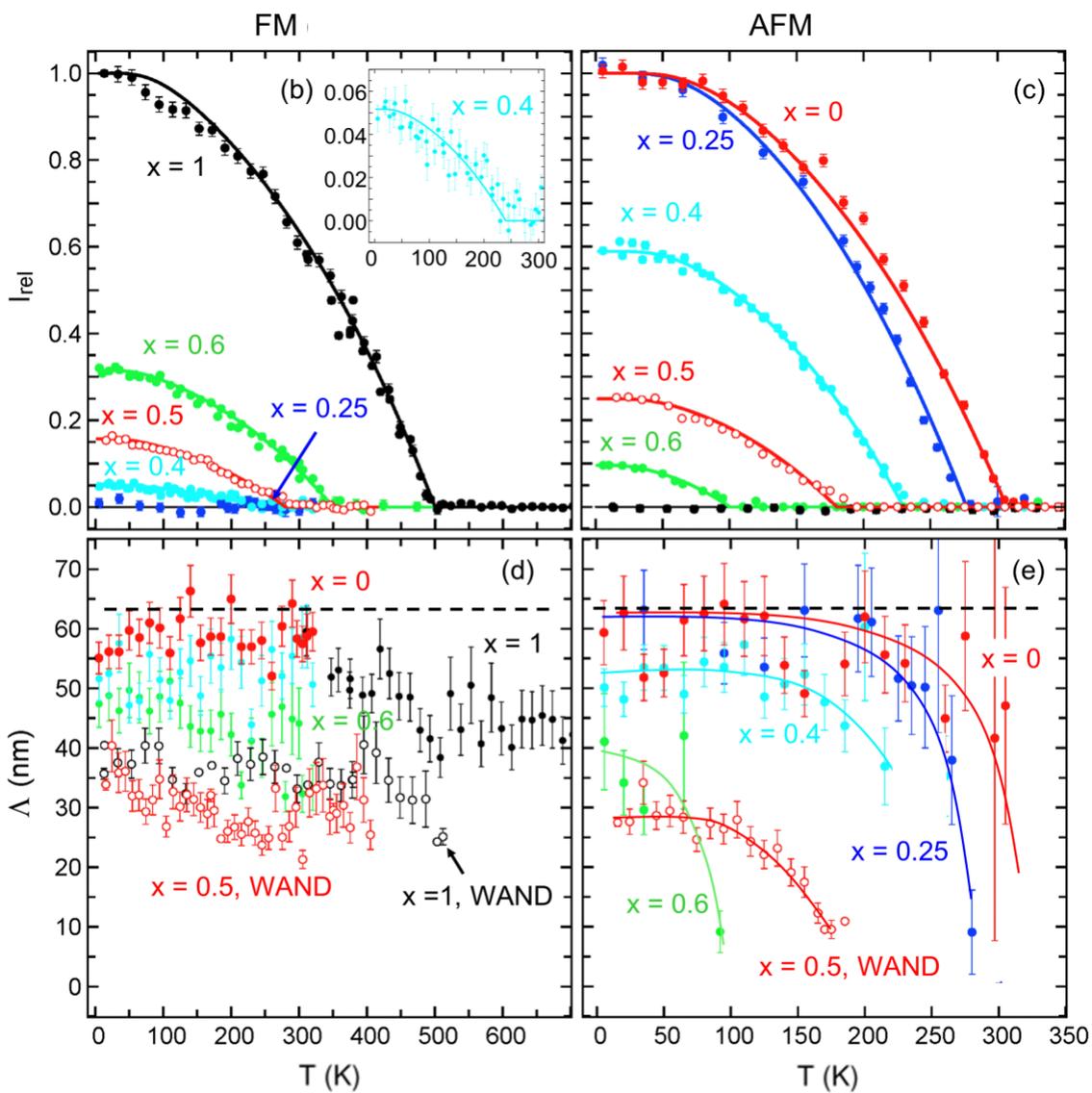

**Figure 2:** (a) Example low scattering wavevector ($Q$) NPD patterns for $x = 0.50$ at 350, 170, and 15 K. As shown in the later panels (and Fig. 1), at this $x$ the FM and AFM order turn on at approximately 300 and 200 K, respectively. Temperature ($T$) dependence of the normalized neutron diffraction intensity ($I_{rel}$), for the FM 1 1 1 (b) and AFM 1/2 -3/2 1/2 (c) reflections for various $x$. The inset in (b) is a blow-up of the $x = 0.40$ data. Solid lines are squared mean-field order parameters using $S = 2$, $L = 2$, $J = 4$ for the FM cases and $S = 5/2$, $L = 0$, $J = 5/2$ for the AFM cases, *i.e.*, atomic values for Fe and Mn, respectively. (d,e) $T$ dependence of the Scherrer length ($\Lambda$) extracted from the FM 1 1 1 (b) and AFM 1/2 -3/2 1/2 (c) reflections for various $x$. The solid lines are guides to the eye. All data were acquired on the HB-1A instrument (FIE-TAX) unless labeled "WAND". The approximate length scale that is practically indistinguishable from the instrumental broadenings is shown as the horizontal dashed line. As noted in the text, the WAND broadening is slightly worse than HB-1A (FIE-TAX) (see, for example, $x = 1.00$ in panel (d)).

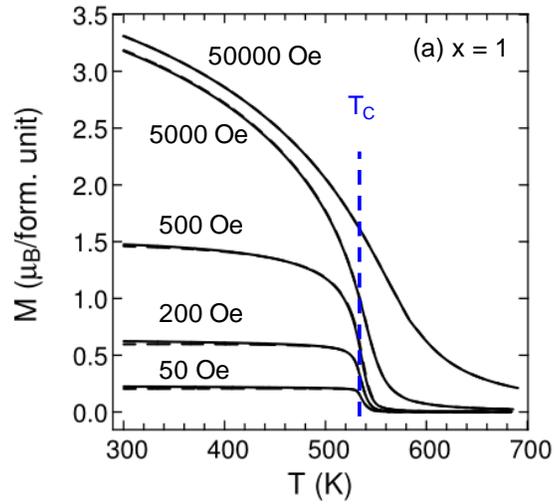

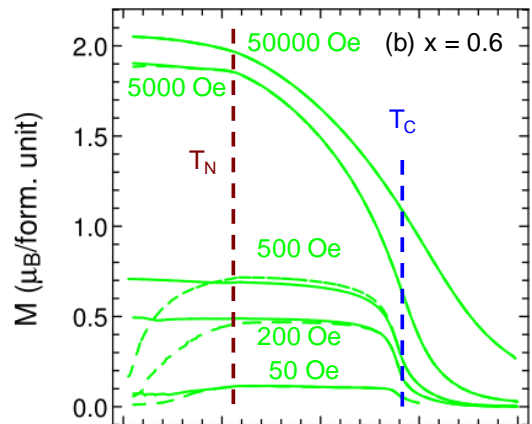

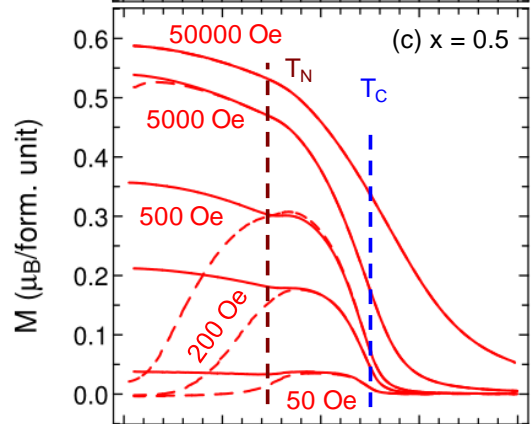

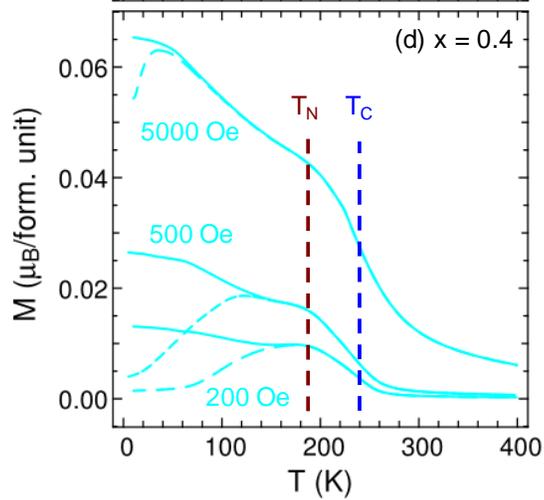

**Figure 3:** Temperature ($T$) dependence of the magnetization ($M$) for (a) $x = 1.00$, (b) $x = 0.60$, (c) $x = 0.50$, and (d) $x = 0.40$, measured in 50, 200, 500, 5000, and 50000 Oe magnetic fields. Solid lines are field-cooled (in the same field used for measurement) and dashed lines are zero-field-cooled. The vertical dashed lines indicate the approximate positions of the Néel temperature ($T_N$, red) and Curie temperature ($T_C$, blue). Some modest differences with the $T_C$ and $T_N$ values from neutron diffraction (Figs. 1 and 2) are attributed to the use of different samples (and very different sample masses) in the two cases, as well as potential high $T$ thermometry issues for large mass neutron samples.

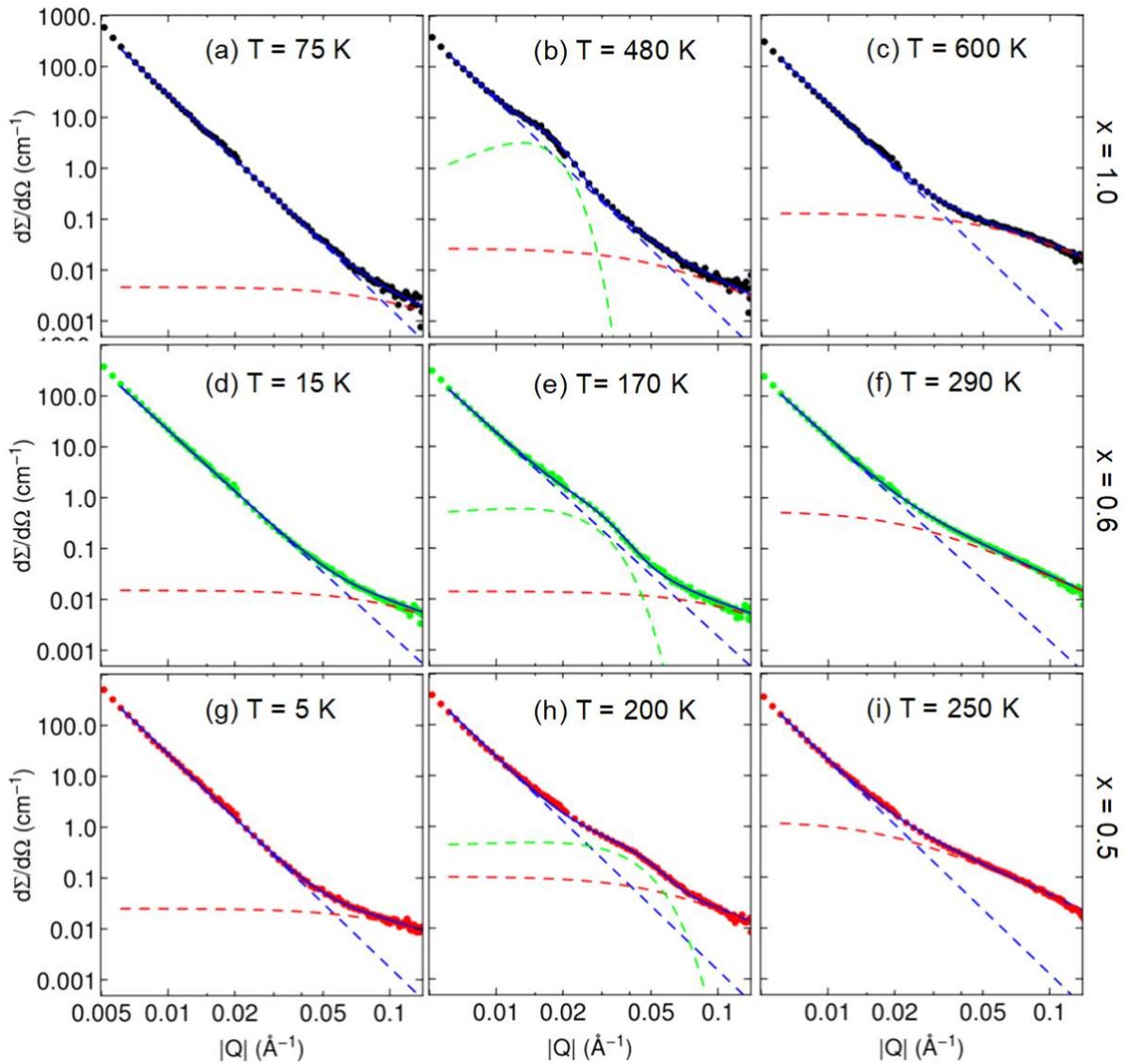

**Figure 4:** SANS cross-section ($d\Sigma/d\Omega$) *vs.* scattering wavevector magnitude ($Q$) for representative $x = 1.00$ (top panels (a-c)), $x = 0.60$ (middle panels (d-f)), and $x = 0.50$ (bottom panels (g-i)) compositions at representative temperatures ($T$). The temperatures decrease from right to left, corresponding to above $T_C$ (the Curie temperature), just below $T_C$, and far below $T_C$, in each case. Solid blue lines are fits based on a sum of Porod (blue dashed lines), Lorentzian (red dashed lines), and Gaussian peak (green dashed lines) contributions.

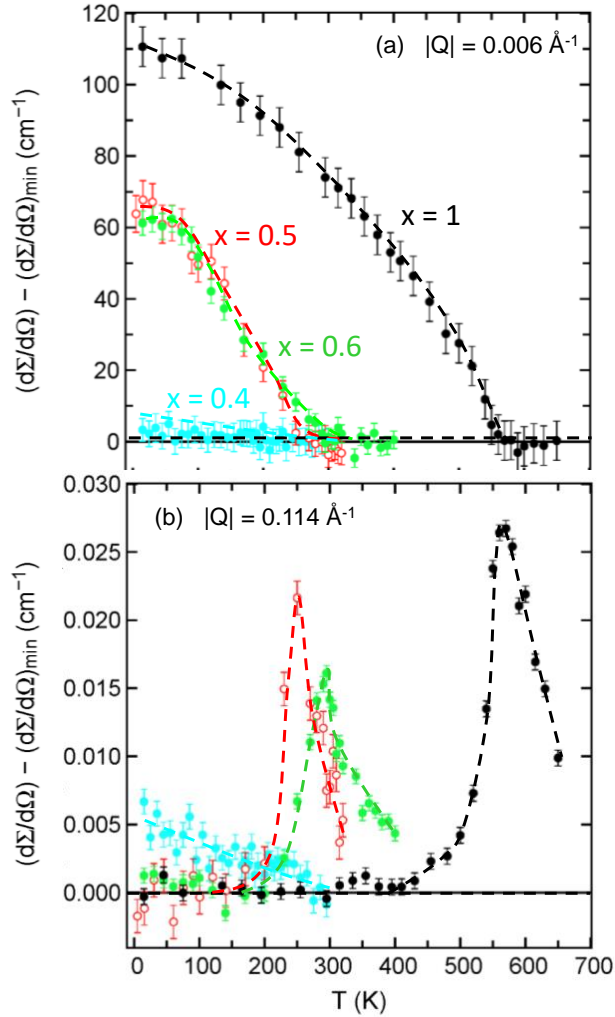

**Figure 5:** SANS cross-section ($d\Sigma/d\Omega$) *vs.* temperature ($T$) at scattering wavevector magnitudes ($Q$) of (a) 0.006 Å$^{-1}$ and (b) 0.114 Å$^{-1}$, for representative $x$ = 1.00, 0.60, 0.50, and 0.40 compositions. Dashed lines are guides to the eye. The data are plotted as [$(d\Sigma/d\Omega) - (d\Sigma/d\Omega)_{min}$], where $(d\Sigma/d\Omega)_{min}$ is the averaged minimum (typically high $T$) value of $d\Sigma/d\Omega$; this highlights the magnetic scattering component. Some modest differences with the $T_C$ values from neutron diffraction (Figs. 1 and 2) are attributed to the use of different samples (and very different sample masses) in the two cases, as well as potential high $T$ thermometry issues for large mass neutron samples.

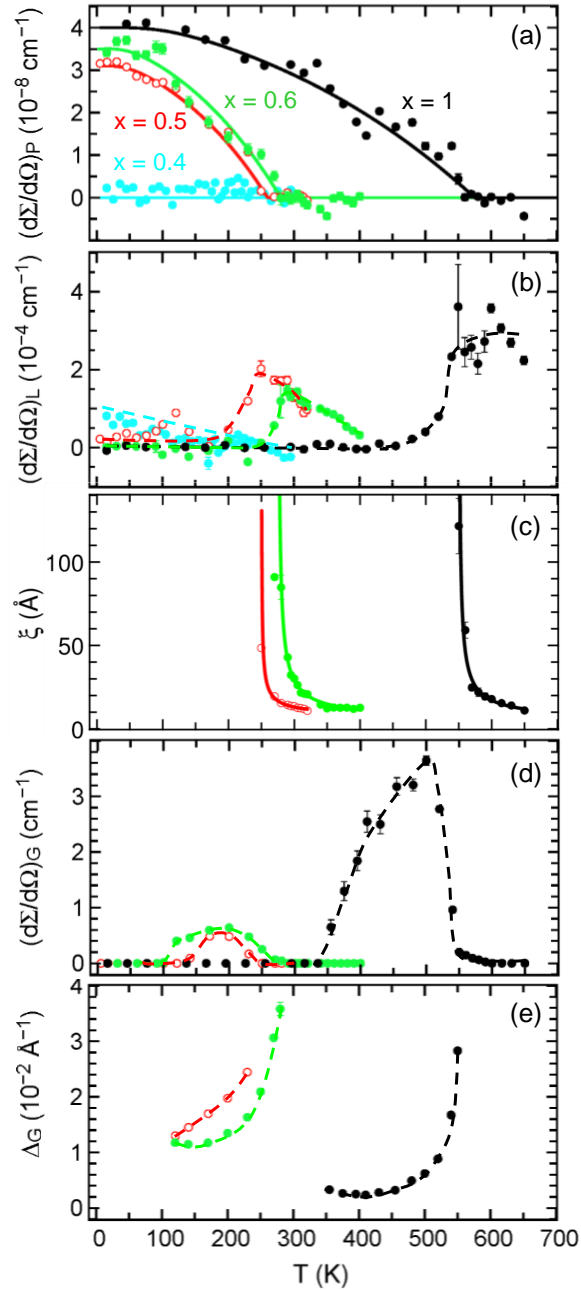

**Figure 6:** Temperature ($T$) dependence of (a) the Porod cross-section ($d\Sigma/d\Omega$)$_\mathrm{P}$, (b) the Lorentzian cross-section ($d\Sigma/d\Omega$)$_\mathrm{L}$, (c) the magnetic correlation length ($\xi$) from the Lorentzian contribution, (d) the Gaussian cross-section ($d\Sigma/d\Omega$)$_\mathrm{G}$, and (e) the Gaussian peak width ($\Delta_\mathrm{G}$), for representative $x = 1.00$, $0.60$, $0.50$, and $0.40$ compositions. Solid lines in (a) are squared mean-field order parameter fits (as in Figs. 2(a,b)), while solid lines in (c) are power law fits, as described in the text. Dashed lines in (b), (d) and (e) are guides to the eye. Some modest differences with the $T_C$ values from neutron diffraction (Figs. 1 and 2) are attributed to the use of different samples (and very different sample masses) in the two cases, as well as potential high $T$ thermometry issues for large mass neutron samples.

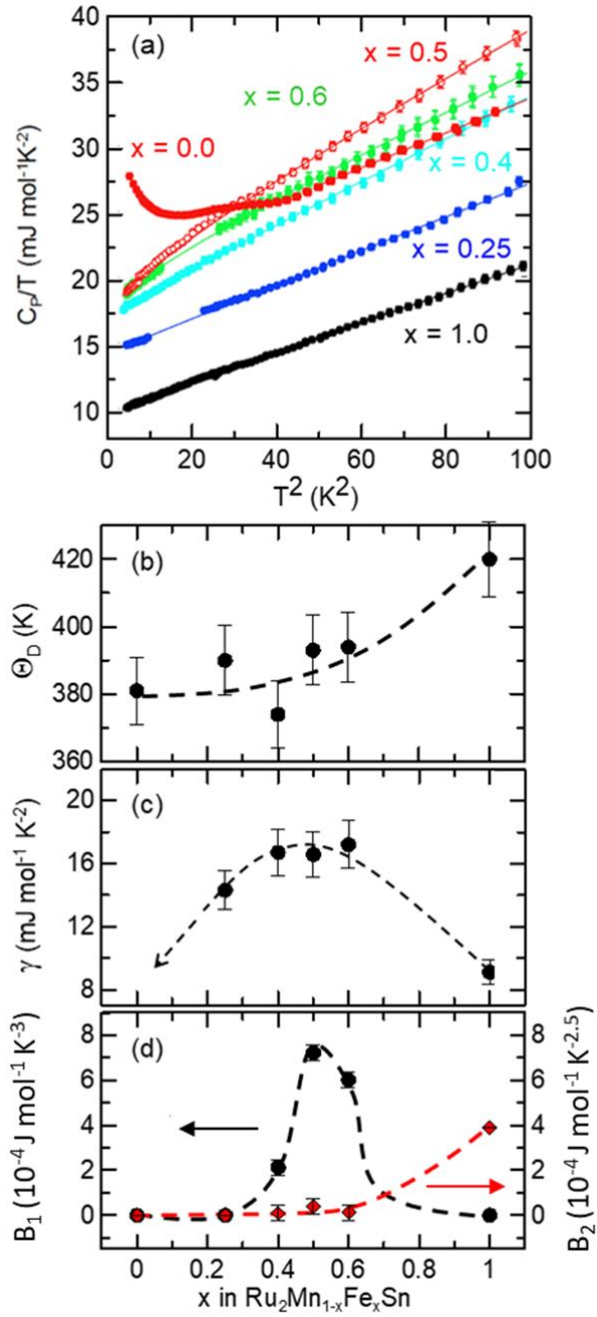

**Figure 7:** (a): Low temperature ($T \leq 10$ K) specific heat ($C_P$) plotted as $C_P/T$ *vs.* $T^2$ for $x$ = 0.00, 0.25, 0.40, 0.50, 0.60 and 1.00. Solid lines are fits to $C_P(T) = \gamma T + \beta T^3 + B_1 T^2 + B_2 T^{3/2}$, as described in the text. $x$ dependence of (b) the Debye temperature ($\Theta_D$), (c) the Sommerfeld coefficient ($\gamma$), and (d) the parameters $B_1$ (left axis) and $B_2$ (right axis). Dashed lines are guides to the eye. No $x$ =

0.00 value of $\gamma$ is shown in (c) due to the Schottky anomaly obscuring the electronic contribution

(see (a)).

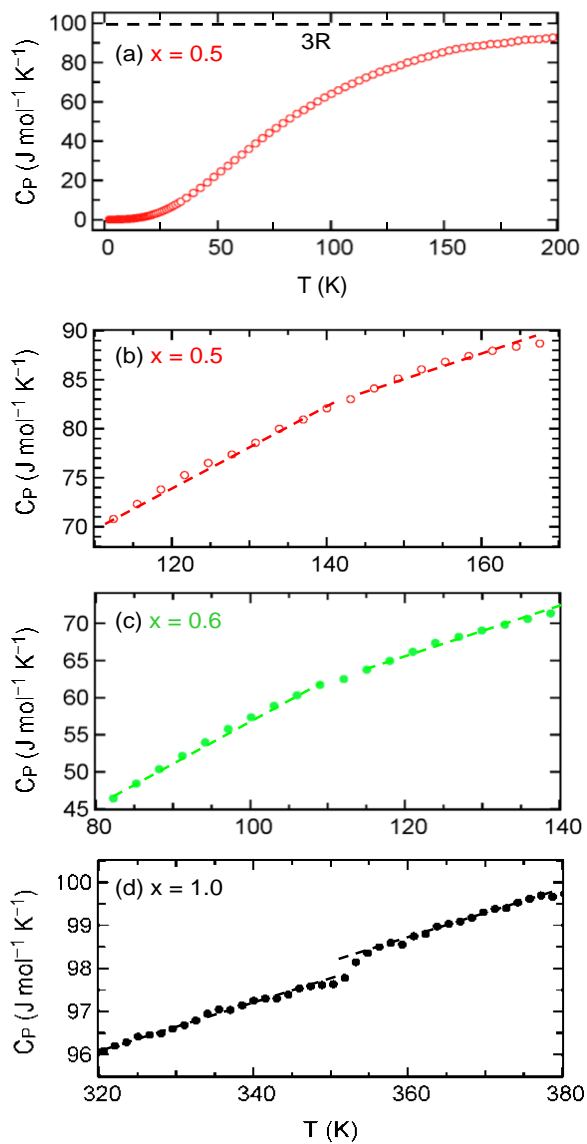

**Figure 8:** (a) Temperature ($T$) dependence of the specific heat ($C_P$) over a wide temperature range (up to 200 K) for a representative $x = 0.50$ composition. The horizontal dashed line marks the Dulong-Petit value ($3R$, where $R$ is the molar gas constant) for reference. $T$ dependence of $C_P$ for $x = 0.50$ (b), $x = 0.60$ (c), and $x = 1.00$ (d) at temperatures around $T*$ (as defined in the text). Dashed lines are guides to the eye. Note the differing $T$ axes in (a-d).

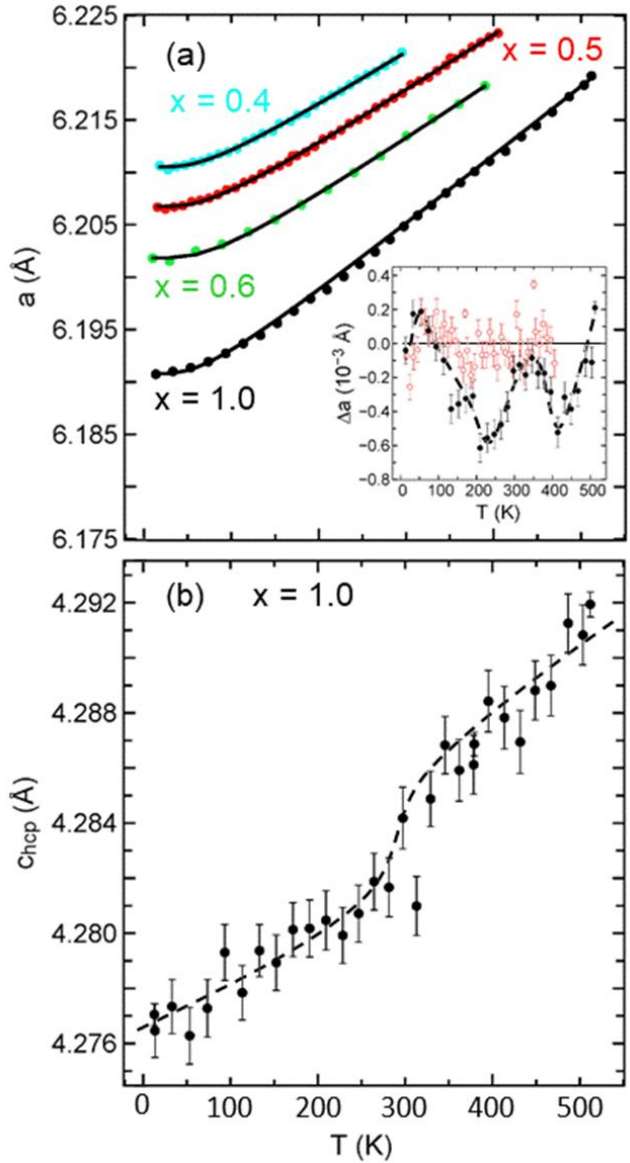

**Figure 9:** (a) Temperature ($T$) dependence of the Heusler cubic lattice parameter ($a$) for representative $x = 0.40$, $0.50$, $0.60$, and $1.00$ compositions. These were determined from Pawley fits to neutron powder diffraction patterns. Solid lines are Grüneisen-Einstein fits, as described in the text. The inset shows the deviation ($\Delta a$) between the data and fits for $x = 0.50$ and $1.00$. (b) $T$ dependence of the $c$-axis lattice parameter ($c_{hcp}$) of the HCP (hexagonal-close-packed) secondary phase for $x = 1$. This was determined from the highest intensity (101) peak, assuming a constant $c/a$ ratio of $1.584$, *i.e.*, that of pure HCP Ru at 300 K.

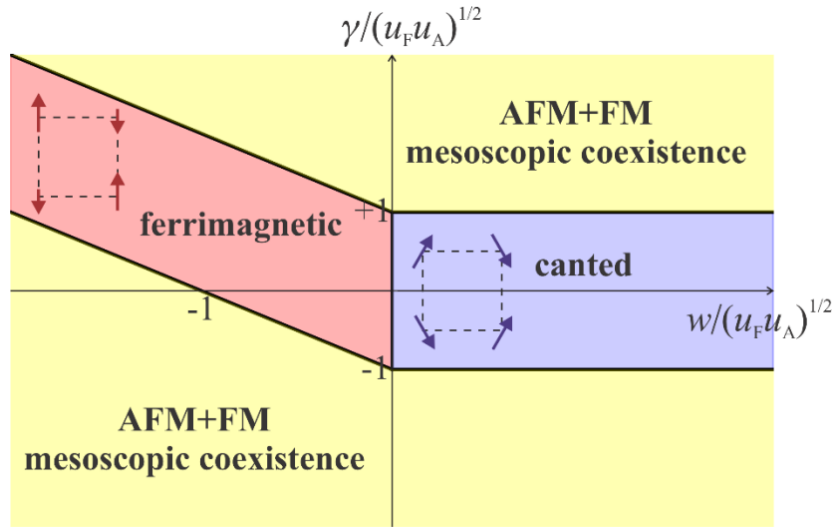

**Figure 10:** Nature of the magnetic state below the multi-critical point of the phenomenological Landau model for competing AFM and FM states. $\gamma$, $w$, $u_F$, and $u_A$ are the quartic coefficients of the free-energy expansion. Depending on their ratios, the system may either form a new magnetic ground state where the AFM and FM order parameters are simultaneously non-zero at every site (red and blue shaded areas) or a state where finite-size regions with only AFM or FM order coexist at the meso-scale (yellow shaded areas). In the former case, the multi-critical point is tetracritical, and there are two types of magnetic ground states depending on the sign of $w$: a ferrimagnetic phase (illustrated in the red shaded area), in which the FM and AFM order parameters are parallel to each other; and a canted spin phase (illustrated in the blue shaded area), where the FM and AFM order parameters are perpendicular. In the case of AFM-FM mesoscopic coexistence, the multi-critical point is bicritical.